# Algebraic bright and vortex solitons in defocusing media

Olga V. Borovkova,[1] Yaroslav V. Kartashov,[1] Boris A. Malomed,[2,1] and Lluis Torner[1]

[1]ICFO-Institut de Ciencies Fotoniques, and Universitat Politecnica de Catalunya, Mediterranean Technology Park, 08860 Castelldefels (Barcelona), Spain

[2]Department of Physical Electronics, School of Electrical Engineering, Faculty of Engineering, Tel Aviv University, Tel Aviv 69978, Israel

We demonstrate that spatially inhomogeneous defocusing nonlinear landscapes with the nonlinearity coefficient growing toward the periphery as $(1+|r|^\alpha)$ support one- and two-dimensional fundamental and higher-order bright solitons, as well as vortex solitons, with algebraically decaying tails. The energy flow of the solitons converges as long as nonlinearity growth rate exceeds the dimensionality, i.e., $\alpha > D$. Fundamental solitons are always stable, while multipoles and vortices are stable if the nonlinearity growth rate is large enough.

Optical solitons may form in many physical settings. By and large, in uniform conservative media, focusing and defocusing nonlinearities give rise to bright and dark solitons, respectively. The situation may change in the presence of transverse modulations of the refractive index (linear lattices), which affect the strength and sign of the effective diffraction for the propagating waves, and may result in the formation of gap solitons even in defocusing media [1,2]. However, guiding bright solitons by the defocusing nonlinearity without the help of a linear potential is commonly considered impossible. At the same time, recent advances in the technology of fabrication of nonlinear materials indicate that not only the refractive index, but also nonlinearity may be profiled, as needed, in the transverse directions. Solitons in such nonlinearity landscapes (nonlinear lattices) may have unusual properties, because the corresponding effective inhomogeneity of the material depends on the intensity of the nonlinear excitation [2]. Many effects were predicted in nonlinear [3-7] and mixed linear-nonlinear [8-14] lattices, in both one- (1D) and two-dimensional (2D) [15-17] settings. However, in contrast to linear lattices, localized and periodic landscapes of defocusing nonlinearities do not support bright solitons (instead, "anti-dark" modes can be built on top of a flat background [18,19]). While a dip in a uniform defocusing background may result in considerable concentration of light, a linear trapping potential is still necessary for the complete localization [20].

In this Letter we show that, in contrast to the common belief, a spatially profiled defocusing nonlinearity whose strength grows toward the periphery does support bright solitons in conservative media. They exist because the growth of the nonlinearity coefficient makes the governing equation non-linearizable for decaying tails, in contrast to media with homogeneous or periodic nonlinearities, where the presence of the decaying tails places soliton into the semi-infinite spectral gap of the linearized system, in which defocusing nonlinearities cannot support localization. We consider an algebraic profile of the nonlinearity coefficient, $\sigma(r) \sim (1+|r|^\alpha)$, that supports bright solitons, multipoles, and vortices in both 1D and 2D geometries. Solitons exist when the nonlinearity growth rate $\alpha$ exceeds certain critical value, and they may be stable. Notice that examples of solitons with algebraic tails in conserv-ative systems are rare [21] (whilst they may exist in defocusing media with nonlinear dissipation [22,23]).

The propagation of light in the nonlinear medium obeys the cubic Schrödinger equation for scaled field amplitude $q$:

$$i\frac{\partial q}{\partial \xi} = -\frac{1}{2}\nabla^2 q + \sigma(\mathbf{r})|q|^2 q, \qquad (1)$$

where $\xi$ is the propagation distance, the transverse coordinates are $\mathbf{r} = \eta$ in the 1D setting and $\mathbf{r} = (\eta, \zeta)$ in 2D setting, $\nabla^2 = \partial^2/\partial\eta^2 + \partial^2/\partial\zeta^2$ accounts for the paraxial diffraction, and $\sigma(\mathbf{r}) > 0$ is the strength of defocusing nonlinearity that varies with $\mathbf{r}$. The transverse variation of the nonlinearity in optical materials can be achieved in different ways [2]. In particular, in photorefractive media, such as $LiNbO_3$, nonuniform doping with Cu or Fe may considerably enhance the local nonlinearity [24]. Equation (1), with $\xi$ replaced by time, also describes the evolution of matter waves in BEC, where the spatially modulated nonlinearity landscape can be induced by means of the Feshbach resonance in nonuniform external fields [2,25].

One of the most important issues addressed in this Letter is the minimal growth rate of the defocusing nonlinearity at $r \to \infty$, which is necessary for the existence of bright solitons. For this purpose, we consider an algebraic modulation profile, $\sigma(\mathbf{r}) = 1 + \kappa|r|^\alpha$, with $\kappa = 1$ fixed by scaling, and vary power $\alpha$. Soliton solutions to Eq. (1) are looked for as $q(r,\xi) = w(r)\exp(im\phi)\exp(ib\xi)$, where $m$ is the topological charge of 2D vortices [in 1D, the term $\exp(im\phi)$ is dropped], $\phi$ the angular coordinate, and $b$ the propagation constant. Examples of 1D and 2D fundamental and higher-order modes, including vortices and 1D multipole with $k = 2$ nodes, that clearly feature bright-soliton shapes despite the defocusing sign of the nonlinearity, are shown in Fig. 1.

In this context, shapes of the fundamental solitons can be accurately predicted by the Thomas-Fermi (TF) approximation, which neglects the diffraction term in Eq. (1) to yield a solution $w_{TF}^2 = -b/(1 + |r|^\alpha)$. This approximation yields asymptotically exact algebraic tails of the solitons of all types, irrespective of the topological charge (the TF profiles of the 2D and 1D fundamental solitons are nearly indistinguishable from their numerical counterparts displayed in Fig. 1). To the best of our knowledge, vortices with algebraically decaying tails have not been reported before in conservative media, although they are known in dissipative settings [23]. In contrast to conventional bright solitons, in our setting the tails' decay rate for solitons does not depend on $b$, being determined solely by the nonlinearity modulation rate $\alpha$. The center of the soliton is always located at the minimum of the defocusing nonlinearity at $r = 0$. The increase of $|b|$ entails the increase of the soliton's amplitude (in agreement with the TF approximation), and a weak contraction of the vortex rings. Further, an expansion of the soliton's wave form near a maximum point $w = w_{max}$ at $r = r_{max}$ (for the fundamental solitons, $r_{max} = 0$), yields $b \leq -(1 + r_{max}^\alpha)w_{max}^2$, hence solitons exist only for $b \leq 0$, as predicted by the TF approximation too.

The soliton families are characterized by the energy flow (norm), $U = \int |q|^2 d\mathbf{r}$. For the fundamental solitons, the TF approximation predicts $U_{TF} = 2\pi^D |b|/[\alpha \sin(\pi D/\alpha)]$, which, in particular, explains the linear dependence $U(b)$ in Fig. 2(a). Vortex solitons in 2D and multipoles in 1D always feature smaller norms than the corresponding fundamental solitons. At fixed $b$ the soliton's energy flow monotonically increases with decrease of $\alpha$ (this is accompanied by progressive soliton expansion in the transverse plane) and diverges when $\alpha$ approaches certain critical value $\alpha_{cr}$ [see Fig. 2(b) for 2D solitons and Fig. 2(c) for 1D solitons]. Importantly, this critical value is directly linked with the number of transverse dimensions $D$ in the problem, i.e. $\alpha_{cr} = D$. This result is in agreement with the prediction of TF approximation where $U_{TF} \to \infty$ at $\alpha \to D$. Thus, steeper defocusing-nonlinearity landscapes are required for the existence of the bright solitons in the higher dimension.

A key issue is the stability of the obtained soliton solutions. In 2D case the linear stability analysis was performed by substituting the perturbed field in the form $q = [w(r) + u(r)\exp(in\phi + \delta\xi) + v^*(r)\exp(-in\phi + \delta^*\xi)]\exp(im\phi + ib\xi)$, with $u, v$ being small perturbations and $n$ being azimuthal perturbation index, into Eq. (1), linearizing it, and solving the eigenvalue problem for perturbation growth rates $\delta$:

$$
\begin{aligned}
i\delta u &= -\frac{1}{2}\left(\frac{d^2}{dr^2} + \frac{1}{r}\frac{d}{dr} - \frac{(m+n)^2}{r^2}\right)u + bu + \sigma w^2(v + 2u), \\
i\delta v &= +\frac{1}{2}\left(\frac{d^2}{dr^2} + \frac{1}{r}\frac{d}{dr} - \frac{(m-n)^2}{r^2}\right)v - bv - \sigma w^2(u + 2v),
\end{aligned}
\qquad (2)
$$

The results of stability analysis for 2D solitons are summarized in Fig. 3. Fundamental solitons are always stable. While vortex solitons with topological charges $m=1$ are weakly unstable at $\alpha \to \alpha_{cr}$ with respect to perturbations with $n=1$, they become stable with increase of $b$. At moderate $\alpha$ values the perturbation growth rate acquires maximum at moderate $b$ value and then it asymptotically vanishes when $b \to -\infty$ [Fig. 3(a)]. Already at $\alpha = 5$ the vortex with $b = -30$ is a completely stable object. The maximal possible perturbation growth rate (determined over all $b$ values) also rapidly diminishes with increase of $\alpha$ [Fig. 3(b)]. Already at $\alpha \sim 8$ the entire family of vortex solitons may be considered stable. A similar picture was encountered for $m=2$ vortex solitons, although such states usually require larger rates $\alpha$ of nonlinearity growth for their stabilization. 1D fundamental solitons are also stable, while stabilization of multipole solitons is achieved for moderate $\alpha \sim 5, 6$ (the larger is the number of azimuthal nodes, the larger is the $\alpha$ value required for stabilization).

The stability analysis was verified by direct simulations of Eq. (1). Figure 4(a) shows a typical decay scenario for an unstable vortex with $m=1$, that develops azimuthal modulations and then transforms into a fundamental soliton, while the phase singularity is expelled to $r \to \infty$. In the unstable vortex with $m=2$, the central singularity may split into two singularities, which is followed by the expulsion of one of them, while the other stays at the center, developing into a stable vortex with $m=1$ [Fig. 4(b)]. Examples of stable propagation of perturbed vortices with charges $m=1$ and $2$ are displayed in Fig. 4(c). The stable modes preserve their internal structure over indefinitely long distances even in the presence of strong input perturbations.

To conclude, we have found that, in contrast to common expectations, spatially inhomogeneous *defocusing* nonlinearities may support stable *bright* solitons with algebraic tails. Along with the fundamental solitons, vortex solitons and one-dimensional multipoles may be also stable. The energy flow of the solitons converges as long as the nonlinearity growth rate in the medium exceeds the dimensionality of the diffracting beam.

# References with titles

# References without titles

# Figure captions

Figure 1. (a) Profiles of 2D solitons with different topological charges corresponding to $b=-10$, $\alpha=5$. (b) Vortices with $m=1$ corresponding to $b=-5$ (curve 1), $b=-10$ (curve 2), and $b=-20$ (curve 3) at $\alpha=5$. (c) 1D solitons with different numbers of nodes corresponding to $b=-10$, $\alpha=5$. The nonlinearity profiles are shown by red curves.

Figure 2. The energy flow versus the propagation constant for 2D solitons at $\alpha=5$ (a), and versus $\alpha$ for 2D (b) and 1D (c) solitons at $b=-10$. Red curves in (b) and (c) show the prediction of the TF approximation.

Figure 3. (a) The real part of the perturbation growth rate versus $b$ at $\alpha=3.5$, 4, and 5 (curves 1, 2, and 3) for vortices with $m=1$ and perturbation index $n=1$. (b) The largest perturbation growth rate for the vortex with $m=1$ versus $\alpha$.

Figure 4. (a) Decay of unstable vortex with $m=1$, $b=-3$, $\alpha=3.5$. (b) Decay of unstable vortex with $m=2$, $b=-20$, $\alpha=4$. (c) Stable propagation of vortex with $m=1$, $b=-30$, $\alpha=5$ (left, center) and vortex with $m=2$, $b=-15$, $\alpha=10$ (right). In all cases, white noise was added into input field distributions.

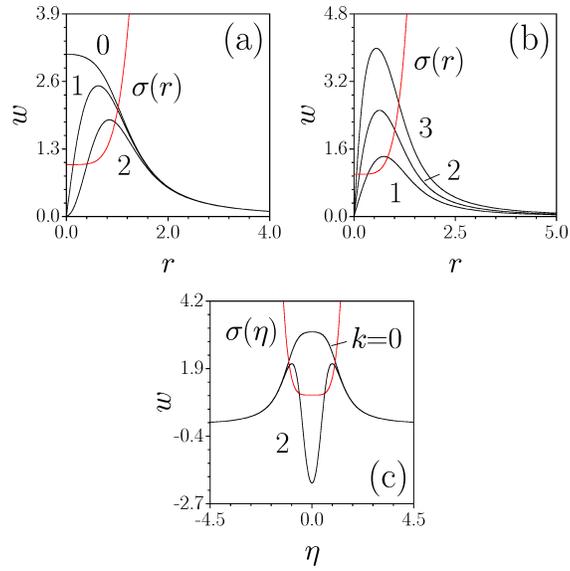

Figure 1. (a) Profiles of 2D solitons with different topological charges corresponding to $b=-10$, $\alpha=5$. (b) Vortices with $m=1$ corresponding to $b=-5$ (curve 1), $b=-10$ (curve 2), and $b=-20$ (curve 3) at $\alpha=5$. (c) 1D solitons with different numbers of nodes corresponding to $b=-10$, $\alpha=5$. The nonlinearity profiles are shown by red curves.
9

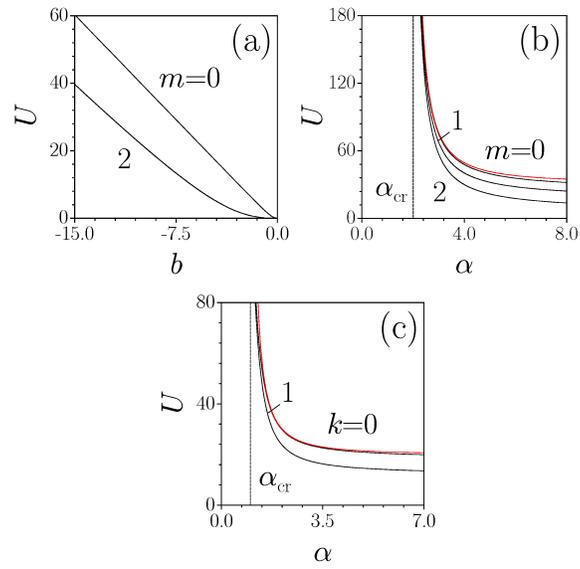

Figure 2. The energy flow versus the propagation constant for 2D solitons at $\alpha=5$ (a), and versus $\alpha$ for 2D (b) and 1D (c) solitons at $b=-10$. Red curves in (b) and (c) show the prediction of the TF approximation.

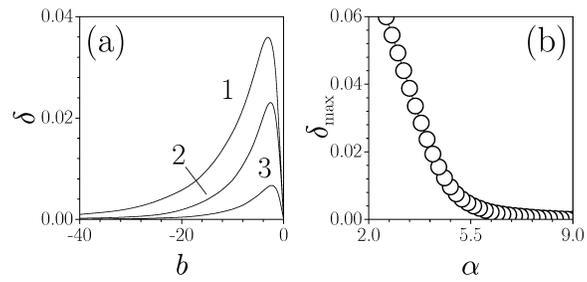

Figure 3.  (a) The real part of the perturbation growth rate versus $b$ at $\alpha = 3.5$, 4, and 5 (curves 1, 2, and 3) for vortices with $m=1$ and perturbation index $n=1$.
(b) The largest perturbation growth rate for the $m=1$ versus $\alpha$.

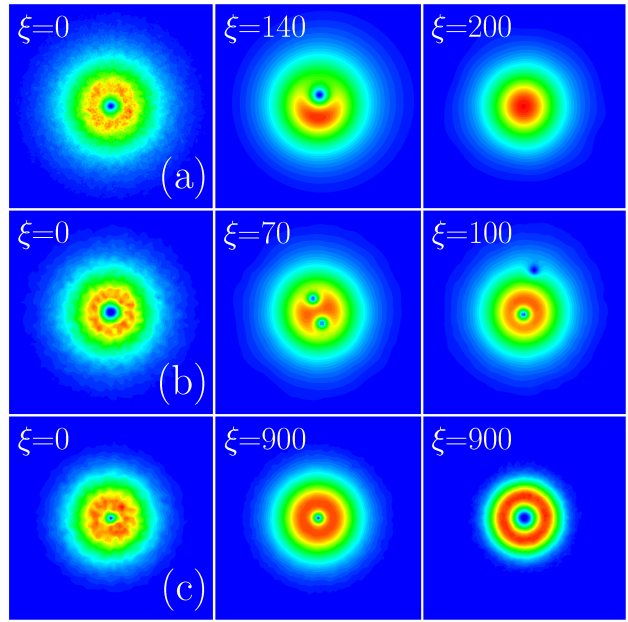

Figure 4. (a) Decay of unstable vortex with $m=1$, $b=-3$, $\alpha=3.5$. (b) Decay of unstable vortex with $m=2$, $b=-20$, $\alpha=4$. (c) Stable propagation of vortex with $m=1$, $b=-30$, $\alpha=5$ (left, center) and vortex with $m=2$, $b=-15$, $\alpha=10$ (right). In all cases, white noise was added into input field distributions.